\documentclass{article}

\usepackage{arxiv}
\usepackage{bbm}
\usepackage[utf8]{inputenc} 
\usepackage[T1]{fontenc}    
\usepackage{url}            
\usepackage{booktabs}       
\usepackage{amsfonts}       
\usepackage{nicefrac}       
\usepackage{microtype}      
\usepackage{subfig}         
\usepackage{lipsum}
\usepackage{amsmath}
\usepackage{graphicx}
\graphicspath{ {./images/} }
\usepackage{booktabs}
\usepackage{multirow}
\usepackage{array}
\usepackage{xcolor}
\usepackage{hyperref}
\title{SCG With Your Phone: Diagnosis of Rhythmic Spectrum Disorders in Field Conditions
}

\author{
 Peter Golenderov, Yaroslav Matushenko, Anastasia Tushina, Michal Barodkin  \\
}

\begin{document}
\maketitle
\begin{abstract}
Aortic valve opening (AO) events are crucial for detecting frequency and rhythm disorders, especially in real-world settings where seismocardiography (SCG) signals collected via consumer smartphones are subject to noise, motion artifacts, and variability caused by device heterogeneity. In this work, we present a robust deep-learning framework for SCG segmentation and rhythm analysis using accelerometer recordings obtained with consumer smartphones. We develop an enhanced U-Net v3 architecture that integrates multi-scale convolutions, residual connections, and attention gates, enabling reliable segmentation of noisy SCG signals. A dedicated post-processing pipeline converts probability masks into precise AO timestamps, whereas a novel adaptive 3D-to-1D projection method ensures robustness to arbitrary smartphone orientation. Experimental results demonstrate that the proposed method achieves consistently high accuracy and robustness across various device types and unsupervised data-collection conditions. Our approach enables practical, low-cost, and automated cardiac-rhythm monitoring using everyday mobile devices, paving the way for scalable, field-deployable cardiovascular assessment and future multimodal diagnostic systems.
\end{abstract}


\section{Introduction}
\label{sec:introduction}

Improvements in sensing, networking, and computing technologies have opened up new opportunities for assessing heart function. In the 21st century, cardiovascular diseases remain among the most significant health challenges, affecting both life expectancy and quality of life \cite{who_cvd}. Rational management and monitoring of individuals with cardiovascular problems can effectively improve the organization of healthcare systems worldwide.

Seismocardiography (SCG) is a technique for measuring vibrations produced by the beating heart using an accelerometer \cite{taebi2019recent}. SCG signals arise from the mechanical processes associated with cardiac activity (such as myocardial contraction, blood momentum changes, valve closure, etc.). The characteristics of these signals can provide useful information correlated with cardiovascular pathologies \cite{zanetti2015ballistocardiography}. Although SCG analysis is complex and its clinical integration remains limited, it is nevertheless considered a relatively simple and noninvasive method for assessing cardiac activity and, consequently, a person's cardiac status. SCG signals can offer a promising alternative for long-term heart-rate monitoring in situations where electrocardiogram (ECG) recordings are unavailable \cite{taebi2019recent}.

Compared with existing cardiac monitoring methods, SCG can provide a cost-effective solution with the added advantage of regular and automated measurements. The assessment of cardiac mechanics is essential for comprehensive evaluation of heart performance in both clinical practice and research and is typically performed using ultrasound or ECG recordings. However, these methods are relatively expensive and require qualified specialists as well as specialized clinical equipment.

Cardiac anomalies may occur irregularly and progress unnoticed, which makes early detection particularly challenging. SCG may additionally enable the assessment of mechanical heart function or the condition of coronary vessels. SCG signals are susceptible to various noise characteristics, which may help reveal different aspects of cardiac mechanical activity. Thus, there is significant interest in implementing SCG-based noninvasive assessment of cardiac function as a tool for continuous heart-status monitoring.

SCG can be assessed repeatedly over long periods of time, supporting periodic measurements and tracking changes in a patient's condition under the influence of various everyday factors. However, most existing SCG assessment systems are designed for laboratory settings or controlled behavioral and environmental conditions, which limits their applicability in real-world scenarios and hampers high-frequency measurements.

To address this limitation, we collected a large-scale dataset comprising more than 7,700 hours of raw SCG recordings obtained from volunteers using built-in accelerometers of consumer smartphones \cite{mehrang2018mobile}. The recordings were captured in diverse real-world conditions, involving different device models and sensor orientations, which naturally increased data variability and realism. These observations represent one of the first large-scale evaluations of SCG outside laboratory environments and demonstrate the feasibility of conducting SCG studies in everyday conditions \cite{mehrang2018mobile}.

However, analyzing SCG signals remains challenging due to their high sensitivity to motion artifacts, sensor placement, and noise \cite{zanetti2015ballistocardiography}. In this study, we propose a deep-learning approach for robust SCG segmentation and rhythm analysis using accelerometer recordings collected from various mobile devices. We demonstrate that combining multi-scale convolutional and attention mechanisms within a modified U-Net architecture \cite{oktay2018attention, li2021integration}, trained with hybrid loss functions \cite{lin2017focal}, along with the proposed pre- and post-processing methods, can significantly improve segmentation stability and performance under noisy conditions.

Experimental results show that the proposed method achieves consistently high accuracy and robustness across various device types and unsupervised data-collection conditions. This enables a transition from limited laboratory research toward full-scale commercial deployment. The practical capabilities of the proposed method are demonstrated on our platform.\footnote{https://heartscan.app}\footnote{https://www.openscg.org/}

\section{Related Work}
\label{sec:related_works}

\subsection{Seismocardiography for Cardiac Monitoring}
Seismocardiography (SCG) is a non-invasive technique that measures chest-wall vibrations caused by cardiac activity, providing insights into mechanical processes such as myocardial contraction and valve motion \cite{taebi2019recent, zanetti2015ballistocardiography}. Unlike traditional methods, such as electrocardiography (ECG) and ultrasound, SCG offers a cost-effective and accessible alternative for long-term cardiac monitoring, particularly in non-clinical settings \cite{taebi2019recent}. Recent studies have explored its potential for detecting cardiovascular anomalies, including atrial fibrillation \cite{mehrang2018mobile, shashikumar2017automated}. However, SCG signals are highly susceptible to motion artifacts and noise, which limits their clinical adoption \cite{zanetti2015ballistocardiography}. Most existing SCG systems are designed for controlled laboratory environments \cite{taebi2019recent}, restricting their applicability in real-world scenarios where environmental and behavioral variability introduces significant challenges.

\subsection{Mobile Devices for SCG Data Collection}
Recent advancements in consumer electronics have enabled the use of smartphone accelerometers for SCG data collection, making it feasible to monitor cardiac activity in everyday conditions \cite{mehrang2018mobile}. For instance, the MODE-AF study demonstrated the potential of smartphone-based SCG for detecting atrial fibrillation in real-world settings \cite{mehrang2018mobile}. Such approaches leverage the widespread availability of smartphones to provide low-cost, scalable solutions for cardiac monitoring. However, the heterogeneity of device specifications and sensor orientations introduces substantial variability, complicating signal analysis \cite{zanetti2015ballistocardiography}. Our work builds on these efforts by collecting a large-scale SCG dataset using consumer smartphones and addressing the challenges associated with real-world data variability.

\subsection{Time Series Analysis and Segmentation}
Traditional SCG analysis relies on time–frequency methods, such as spectrograms or heuristic peak-detection algorithms, to identify key cardiac events \cite{shashikumar2017automated}. However, these methods struggle with noisy data and fail to generalize to real-world conditions. Recent advancements in deep learning have shown considerable promise for analyzing medical time series, including ECG and SCG signals \cite{shen2020deep}. For example, convolutional neural networks (CNNs) have been successfully applied to ECG segmentation, achieving robust performance in identifying cardiac cycles \cite{shen2020deep}. Similarly, U-Net-based architectures, originally developed for image segmentation, have been adapted for time series tasks and demonstrated high accuracy in applications such as sleep-stage classification \cite{perslev2019u}. Our approach extends these methods by applying a modified U-Net architecture for SCG segmentation, addressing the unique challenges posed by noisy, real-world data.

\subsection{Deep Learning for Anomaly Detection and Segmentation}
Deep learning techniques, including autoencoders and transformer-based models, have been explored for anomaly detection in time series data \cite{lee2023anomalybert}. In preliminary tests, we evaluated the Anomaly-BERT model for detecting cardiac events as anomalies, but its performance was limited due to the complexity and variability of SCG signals \cite{lee2023anomalybert}. This motivated a shift toward segmentation-based approaches using encoder–decoder architectures, which have shown promise in handling noisy medical signals \cite{perslev2019u}. Furthermore, attention mechanisms have been integrated into CNNs to improve robustness by enabling models to focus on relevant features \cite{oktay2018attention, li2021integration}. Our proposed method combines multi-scale convolutions with attention mechanisms, inspired by recent work on hybrid architectures \cite{li2021integration}, and employs hybrid loss functions such as Focal Loss to address data imbalance and noise \cite{lin2017focal}.

\subsection{Gaps and Contributions}
While significant progress has been made in SCG research, most studies focus on controlled environments, limiting their practical applicability \cite{taebi2019recent}. Moreover, existing deep learning methods for SCG analysis are often trained on clean data and struggle with real-world noise and device variability \cite{zanetti2015ballistocardiography}. Our work addresses these gaps by developing a robust deep-learning framework for SCG segmentation and rhythm analysis, leveraging a large-scale dataset collected from consumer smartphones. By combining advanced architectural designs with improved loss functions, we aim to enable reliable cardiac monitoring in diverse, uncontrolled real-world settings.

\section{Methods}
\label{sec:methods}
\subsection{Problem Definition}
\label{subsec:problem_definition}
Seismocardiography (SCG) is a non-invasive technique for assessing cardiac mechanics by measuring chest-wall vibrations induced by heart activity -- such as myocardial contractions and valve movements -- using accelerometers \cite{taebi2019recent}. These vibrations manifest as characteristic peaks in SCG signals, with aortic valve opening (AO) points being particularly critical for analyzing heart rate, rhythm, and the sequence of myocardial contractions. Accurate identification of AO peaks is essential for detecting frequency and rhythm disorders, especially in real-world settings where SCG signals collected via consumer mobile devices are subject to noise, motion artifacts, and variability due to device heterogeneity.

Physiologically, the mechanical impulse generated at AO propagates orthogonally to the chest surface (along the dorsoventral direction). Importantly, the temporal morphology of the AO complex is invariant to sensor and body rotation: the waveform pattern is preserved across all axes, differing only in amplitude and polarity due to projection onto the device-specific coordinate system.

In this study, SCG signals were recorded using the built-in accelerometers of consumer smartphones, capturing acceleration data along all three axes $j \in \{x, y, z\}$. Formally, given a time-series SCG signal $\mathbf{a}_j = \{a_j(t)\}_{t=1}^T \in \mathbb{R}^T$, representing acceleration measured along axis $j$ over time interval $T$, the model predicts a probability mask $\hat{y} \in [0,1]^T$, where $\hat{y}(t)$ denotes the likelihood that time $t$ corresponds to an AO event. The model is trained in a supervised fashion using binary ground-truth masks $y \in \{0,1\}^T$. 

Although the binary mask explicitly localizes AO events within a temporal window, the ultimate clinical objective requires identification of the precise AO moment. Extracting these precise timestamps from the probabilistic output is accomplished through a dedicated post-processing stage, described in detail in Subsection~\ref{subsec:postprocessing}.

The goal of this work is to develop a robust deep-learning model capable of accurately segmenting SCG signals to identify AO peaks, enabling reliable analysis of heart rate and rhythm under noisy, real-world conditions. This approach aims to facilitate low-cost, automated, and non-invasive cardiac monitoring using widely available consumer devices.

\subsection{Data and Pre-Processing}
\label{subsec:data_preprocessing}

\begin{figure}[!t]
    \centering
    \subfloat[Distribution of mobile device vendors in the dataset.]{
        \includegraphics[width=0.4\linewidth]{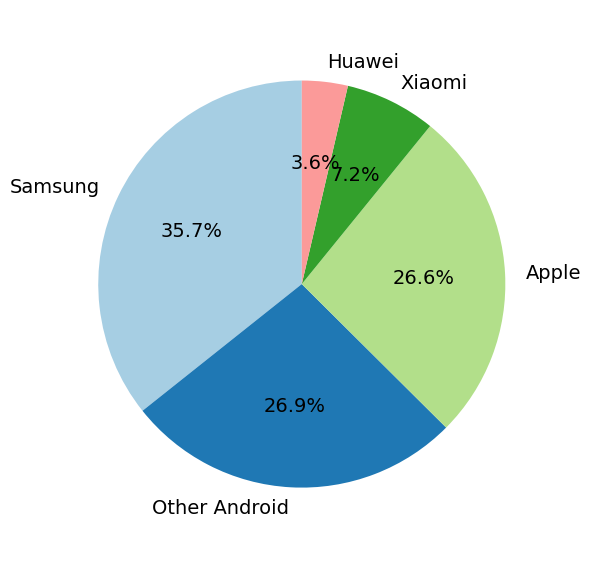}
    }
    \hfill
    \subfloat[Histogram of accelerometer sampling frequencies across the dataset.]{
        \includegraphics[width=0.5\linewidth]{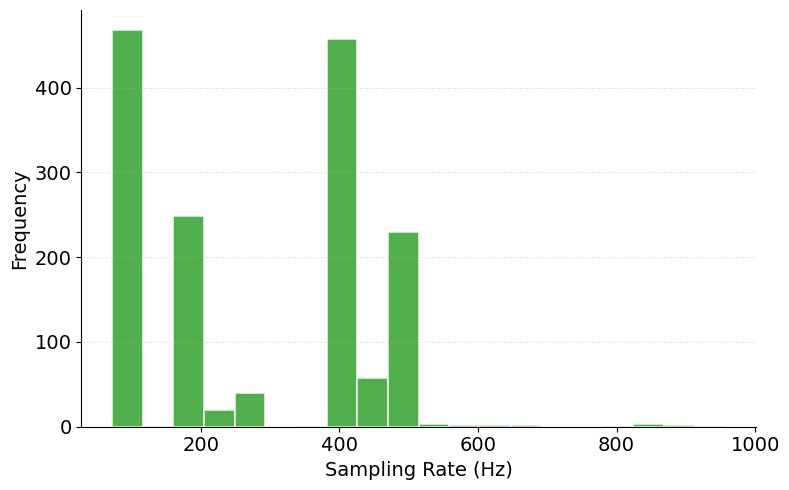}
    }
    \caption{Overview of the dataset statistics: (a) distribution of mobile device vendors and (b) distribution of accelerometer sampling frequencies.}
    \label{fig:dataset_distribution}
\end{figure}

To develop and evaluate our SCG segmentation model, we collected a large-scale dataset comprising over 7700 hours of raw seismocardiography (SCG) recordings from volunteers using the built-in accelerometers of consumer smartphones \cite{mehrang2018mobile}. The recordings were acquired under diverse real-world conditions, including various device models, sensor orientations, and environmental settings, thereby introducing substantial variability and realism into the data. Due to limited resources and the complexity of annotation, cardiac specialists labeled 1800 segments (totaling 3 hours) of representative SCG recordings. An overview of the annotated dataset statistics is presented in Figure~\ref{fig:dataset_distribution}.

\begin{figure}[!t]
  \centering
  \includegraphics[width=0.9\linewidth]{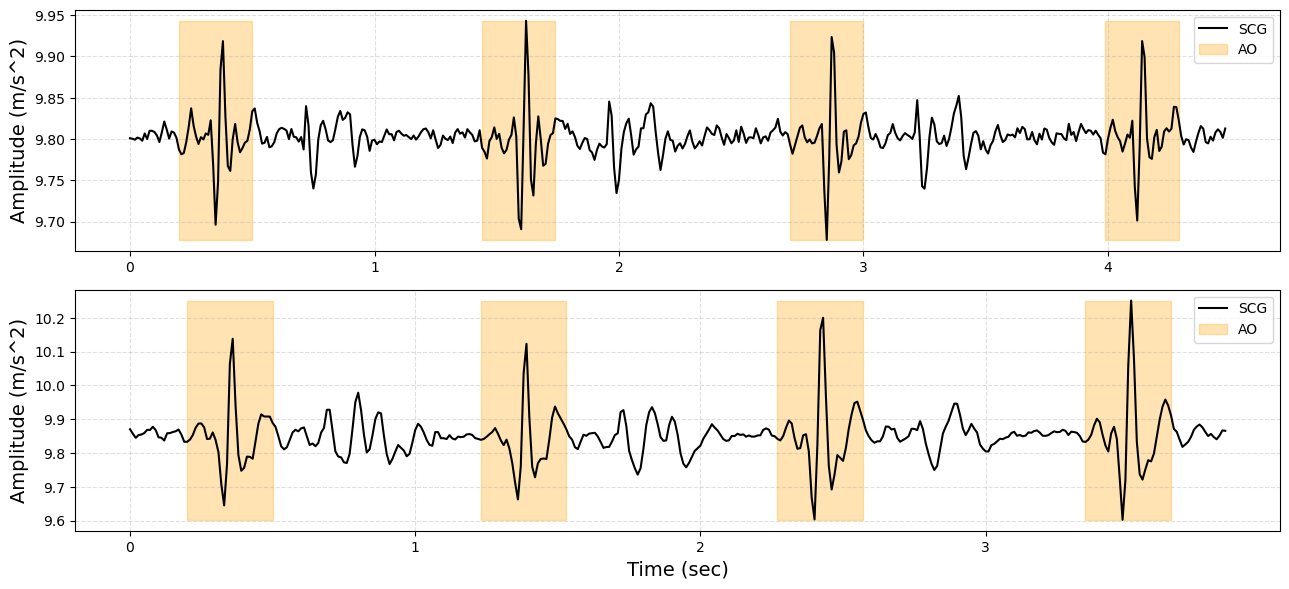}
  \caption{Labeled SCG.}
  \label{fig:labeled_scg}
\end{figure}

The input data are defined as time-series signals $\mathbf{a}_j = \{a_j(t)\}_{t=1}^T \in \mathbb{R}^T$ for segments ranging from 2 to 10 seconds in duration, where $j \in \{x, y, z\}$. The corresponding ground-truth annotations are binary masks $y \in \{0,1\}^T$, where $y(t) = 1$ indicates the presence of an AO peak and $y(t) = 0$ denotes non-peak regions.

The annotation process involved expert labeling of AO peaks on the most informative axis, identifying 2--6 peaks per cardiac cycle based on characteristic temporal patterns. To construct the binary mask $y$, a temporal window centered around each manually annotated peak was applied: values within the window were set to $y(t) = 1$, and $y(t) = 0$ elsewhere. The window half-width was selected empirically. An example of labeled SCG segments is shown in Figure~\ref{fig:labeled_scg}.

To mitigate device-induced variability, all recordings were resampled to a unified sampling rate of 100~Hz using an FFT-based resampling method. Each segment was then independently MinMax-normalized to the range $[-1, 1]$, ensuring scale invariance across the diverse accelerometer sensitivities present in consumer hardware.

To enhance model robustness to temporal variations and improve generalization, data augmentation was applied. Specifically, each segment and its corresponding binary mask were randomly shifted by up to 10\% of the segment duration.

This preprocessing pipeline transforms heterogeneous raw recordings into a standardized, high-quality dataset suitable for deep segmentation. The resulting annotated segments form the foundation for all subsequent experiments.

\subsection{Model Architecture}
\label{subsec:model_architecture}
To address the challenge of segmenting SCG signals for accurate detection of aortic valve opening (AO) peaks in noisy, real-world conditions, we explored two primary classes of neural network architectures: convolutional U-Net-based models and transformer-based models. These architectures were selected based on their proven effectiveness in time-series segmentation tasks \cite{perslev2019u} and on preliminary experiments indicating the limitations of anomaly-detection approaches -- such as Anomaly-BERT \cite{lee2023anomalybert} -- for SCG analysis. The following sections describe the baseline U-Net, its modifications resulting in U-Net v3, and transformer-based adaptations, all evaluated for their ability to robustly segment SCG signals.

\paragraph{Baseline U-Net Architecture.} 
The U-Net architecture, originally developed for medical image segmentation \cite{ronneberger2015unet}, was chosen as the baseline due to its demonstrated success in segmenting medical time-series data, including sleep staging \cite{perslev2019u}. The standard U-Net consists of an encoder–decoder structure with convolutional layers and skip connections that preserve temporal information across layers. The encoder progressively downsamples the input signal to capture higher-level representations, while the decoder upsamples these representations and combines them with skip connections to refine segmentation boundaries. 

However, the baseline U-Net showed limited effectiveness in handling the complex dynamics and substantial noise present in real-world SCG signals. In particular, it struggled to capture rhythmic structures across multiple temporal scales and to maintain robustness under varying noise patterns. These limitations motivated the development of architectural modifications to enhance performance, as detailed in the following section.

\begin{figure}[!t]
    \centering
    \includegraphics[width=1.0\linewidth]{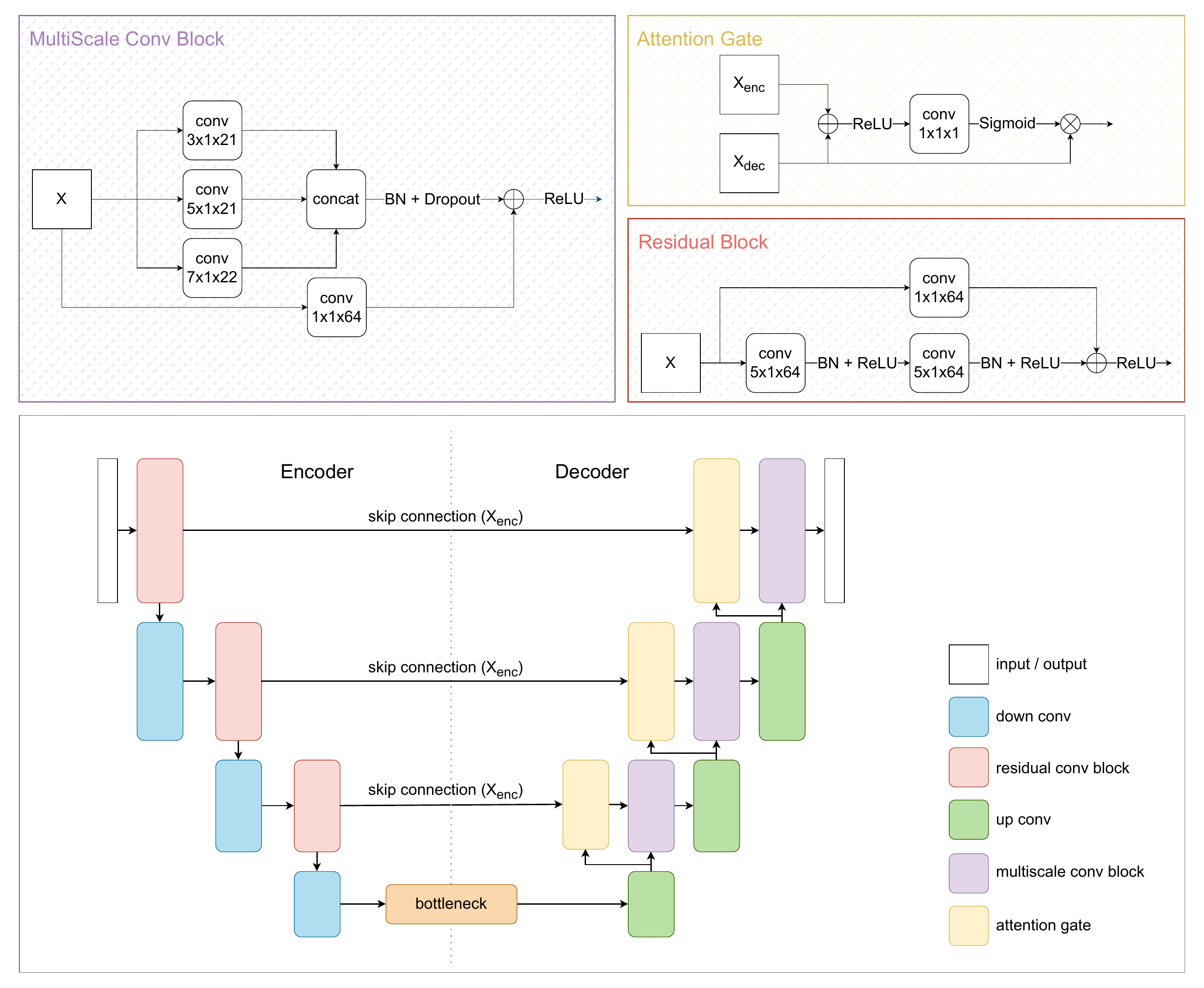}
    \caption{Modified U-Net v3 Architecture.}
    \label{fig:unetv3}
\end{figure}

\paragraph{Modified U-Net Architecture (U-Net v3).} 
To address the shortcomings of the baseline U-Net, we introduced three key modifications that resulted in the U-Net v3 architecture (Figure~\ref{fig:unetv3}), tailored specifically for SCG segmentation. First, multi-scale convolution blocks were incorporated into the encoder to capture both local and global temporal features. These blocks apply parallel convolutional filters with varying kernel sizes, enabling the model to process rhythmic patterns of different lengths and frequencies -- an essential capability for SCG signals characterized by diverse timescales. Second, residual connections were added to stabilize gradient flow during training and preserve information from earlier layers, thereby improving segmentation boundary accuracy. Third, attention gates \cite{oktay2018attention} were integrated at the ends of residual blocks within the skip connections to focus the model on informative signal regions, such as AO peaks, while suppressing noise and irrelevant artifacts. Collectively, these modifications significantly enhanced the model’s robustness to real-world noise and variability, yielding superior performance, as reported in Table~\ref{tab:results}. 

Furthermore, the integration of attention gates improved model interpretability by highlighting physiologically meaningful regions of the signal.

\paragraph{Transformer and Hybrid Architectures.} 
Transformer-based models, widely recognized for their effectiveness in sequential data processing \cite{vaswani2017attention}, were also explored due to their potential in time-series segmentation tasks. The baseline transformer employed a standard self-attention mechanism with fixed sinusoidal positional encoding \cite{vaswani2017attention}. However, this approach struggled to segment SCG signals containing rare and irregular patterns -- such as AO peaks -- largely because the fixed positional encoding was insufficient for capturing the temporal significance of short-duration cardiovascular events. To mitigate this limitation, we introduced learnable positional encoding, allowing the model to adaptively emphasize relevant temporal regions and improving segmentation performance. Additionally, we developed a hybrid architecture by incorporating Self-Attention Convolutional Transformer (SACT) blocks \cite{li2021integration, lin2020conformer}, which combine self-attention for modeling long-range dependencies with convolutional layers for capturing local structures. This hybrid model, referred to as SACT, reduced computational overhead, accelerated training, and further improved segmentation accuracy. Nevertheless, despite these enhancements, transformer-based and hybrid models underperformed compared with U-Net v3, primarily due to the limited size of the annotated dataset and the intrinsic complexity of SCG signal morphology, as shown in Table~\ref{tab:results}.

\paragraph{Summary.} 
The progression from the baseline U-Net to the modified U-Net v3 and transformer-based architectures reflects a systematic effort to address the challenges of SCG segmentation under noisy, real-world conditions. U-Net v3, augmented with multi-scale convolutions, residual connections, and attention gates, achieved the best balance between accuracy and robustness. Its enhanced interpretability -- supported by attention gates and Grad-CAM visualizations \cite{selvaraju2016gradcam} -- further underscores its suitability for clinical and mobile-health applications. Transformer-based models, although promising, were ultimately less effective given the dataset constraints and the complex temporal characteristics of SCG signals.

\subsection{Post-Processing}
\label{subsec:postprocessing}

While the model outputs a probability map $\hat{y}(t) \in [0,1]^T$, the main goal is to extract precise AO timestamps. Therefore, we developed a dedicated post-processing pipeline.

First, the probability map is binarized using a fixed threshold $\tau$, optimized on the validation set:

\[
\hat{y}_{\text{bin}}(t) = \mathbb{1}\{\hat{y}(t) > \tau\}.
\]

To robustly handle fragmented predictions caused by noise, we apply the following steps:

\begin{enumerate}
    \item \textbf{Merge nearby segments:} Small gaps are closed using binary morphological closing, reconnecting AO peak fragments that were split by transient dropout.
    \item \textbf{Remove false-positive detections:} Short isolated segments -- typically arising from motion artifacts or respiration -- are removed as likely false positives.
\end{enumerate}

AO peaks $t_{AO}^*$ are then defined as local maxima within each merged segment $S_i$:

\[
t_{AO}^* = \arg\max_{t \in S_i} |s(t)|, \quad S_i \in S.
\]

\subsection{Loss Functions and Metrics}
\label{subsec:loss_functions_metrics}

\paragraph{Metrics.}  
For segmentation evaluation, we use Intersection over Union (IoU) and the Dice Coefficient to assess the model’s ability to accurately delineate AO peak regions in the output binary mask. These metrics quantify the overlap between predicted and ground-truth masks:

\begin{equation}
\label{eq:iou}
\text{IoU} = \frac{\text{TP}}{\text{TP}+\text{FP}+\text{FN}}.
\end{equation}

\begin{equation}
\label{eq:dice}
\text{Dice} = \frac{2 \times \text{TP}}{( \text{TP}+\text{FP} ) + ( \text{TP}+\text{FN} ) }.
\end{equation}

Since the ultimate objective is the precise detection of the AO peak instant in each cardiac cycle, conventional segmentation metrics alone do not fully capture clinically meaningful performance. Therefore, we additionally compute event-based metrics for AO peaks obtained after the post-processing stage (Subsection~\ref{subsec:postprocessing}). For this task, TP, FP, and FN are defined as follows:

\begin{itemize}
    \item A predicted peak is classified as a \textbf{True Positive (TP)} if it lies within the corresponding ground-truth mask.
    \item A predicted peak with no ground-truth mask is classified as a \textbf{False Positive (FP)}.
    \item A ground-truth mask containing no predicted peak is classified as a \textbf{False Negative (FN)}.
    \item \textbf{True Negatives (TN)} are not defined, since the absence of a peak at an arbitrary time point has no clinical interpretation.
\end{itemize}

Using the counts of TP, FP, and FN, the following event-based metrics are computed across the entire test set:

\begin{equation}
\label{eq:sensitivity}
\text{Sensitivity} = \frac{\text{TP}}{\text{TP} + \text{FN}},
\end{equation}

\begin{equation}
\label{eq:specificity}
\text{PPV} = \frac{\text{TP}}{\text{TP} + \text{FP}},
\end{equation}

\begin{equation}
\label{eq:f1}
\text{F1} = \frac{2 \times \text{PPV} \times \text{Sensitivity}}{\text{PPV} + \text{Sensitivity}}.
\end{equation}

To evaluate post-processing accuracy, we compute the Root Mean Square Error (RMSE):

\begin{equation}
\label{eq:rmse}
\text{RMSE} = \sqrt{\frac{1}{N} \sum_{i=1}^N ( t_{AO,i} - \hat{t}_{AO,i} )^2 },
\end{equation}

where $t_{AO,i}$ and $\hat{t}_{AO,i}$ are the ground-truth and predicted AO timestamps, respectively, and $N$ is the number of AO events.

Together, these metrics provide a comprehensive evaluation of both segmentation quality and precise AO peak detection.

\paragraph{Loss Functions.}  
We tested the following loss functions to optimize the models:

\begin{itemize}
    \item \textbf{Binary Cross-Entropy (BCE)}:
    \begin{equation}
    \text{BCE} = -\frac{1}{T} \sum_{t=1}^T \left[ y(t) \log(\hat{y}(t)) + (1 - y(t)) \log(1 - \hat{y}(t)) \right],
    \label{eq:bce}
    \end{equation}
    which penalizes misclassifications at each time step in the binary mask.
    
    \item \textbf{Dice Coefficient Loss}:
    \begin{equation}
    \text{Dice Loss} = 1 - \frac{2 \sum_{t=1}^T y(t) \cdot \hat{y}(t) + \epsilon}{\sum_{t=1}^T y(t) + \sum_{t=1}^T \hat{y}(t) + \epsilon},
    \label{eq:dice_loss}
    \end{equation}
    where \( \epsilon = 10^{-6} \) prevents division by zero. This loss maximizes segment overlap and reduces false positives.
    
    \item \textbf{Focal Loss} \cite{lin2017focal}:
    \begin{equation}
    \text{Focal Loss} = -\frac{1}{T} \sum_{t=1}^T \left[ y(t) (1 - \hat{y}(t))^\gamma \log(\hat{y}(t)) + (1 - y(t)) \hat{y}(t)^\gamma \log(1 - \hat{y}(t)) \right],
    \label{eq:focal_loss}
    \end{equation}
    with \( \gamma = 2 \), emphasizing hard-to-classify samples such as rare AO peaks.
    
    \item \textbf{Combinations}: We evaluated \textbf{Dice + BCE} (\( \lambda_1 \text{Dice Loss} + \lambda_2 \text{BCE} \)) and \textbf{Dice + Focal} (\( \lambda_1 \text{Dice Loss} + \lambda_2 \text{Focal Loss} \)), with coefficients \( \lambda_1 \) and \( \lambda_2 \) manually tuned during training to balance segmentation precision and stability.
\end{itemize}

\subsection{Adaptive 3D-to-1D Projection}

While the segmentation model is trained on 1D recordings aligned with cardiac activity (Section \ref{subsec:data_preprocessing}), deployment on arbitrary consumer smartphones introduces a critical challenge: device orientation is unknown and may not align with the dorsoventral chest direction. As discussed in Section \ref{subsec:problem_definition}, although the temporal morphology of AO peaks is invariant to sensor rotation, their amplitude depends on misalignment. Excessive misalignment can lead to missed peaks, higher RMSE, and false rhythm detections.

To achieve orientation-agnostic performance, we developed an adaptive 3D-to-1D projection pipeline that estimates the optimal viewing direction $\mathbf{n} \in \mathbb{R}^3$ -- the unit vector aligned with cardiac-induced chest displacement -- using only the raw triaxial signal $\mathbf{a} = (a_x, a_y, a_z)$ and the pretrained U-Net v3. This unsupervised pipeline consists of three stages and introduces no additional learnable parameters.

\paragraph{Axis Selection.}  
The U-Net v3 model is applied independently to each axis, producing predicted AO timestamps and a confidence score $c \in [0,1]$. The axis with the highest confidence is selected as the reference axis:
\[
j^* = \arg\max_{j \in \{x,y,z\}} c_j.
\]

\paragraph{Estimation of Local Oscillation Direction.}  
For each detected AO peak $p_i$ on the reference axis $j^*$, a symmetric temporal window is extracted. Within this window, the global maximum and minimum of the reference signal are identified at time indices $t_{\max}$ and $t_{\min}$. The corresponding 3D acceleration vectors are retrieved:
\[
\mathbf{v}_{\max} = \begin{bmatrix} a_x(t_{\max}) \\ a_y(t_{\max}) \\ a_z(t_{\max}) \end{bmatrix}, \quad
\mathbf{v}_{\min} = \begin{bmatrix} a_x(t_{\min}) \\ a_y(t_{\min}) \\ a_z(t_{\min}) \end{bmatrix}.
\]

The local oscillation vector is computed as:
\[
\mathbf{d}_i = \mathbf{v}_{\max} - \mathbf{v}_{\min},
\]
and normalized:
\[
\hat{\mathbf{d}}_i = \frac{\mathbf{d}_i}{\|\mathbf{d}_i\|_2}.
\]

A robust mean direction is estimated by averaging the retained vectors after discarding outliers deviating more than $45^\circ$ from the mean. If fewer than 3 vectors remain, the fallback direction $\mathbf{e}_z = [0, 0, 1]^\top$ is used. The final oscillation direction is:
\[
\mathbf{n} = \frac{1}{K} \sum_{k=1}^K \mathbf{d}_k, \quad \mathbf{n} \leftarrow \frac{\mathbf{n}}{\|\mathbf{n}\|_2},
\]
where $K$ is the number of retained vectors.

\paragraph{Projection and Final Detection.}  
Each axis is zero-mean centered:
\[
\tilde{a}_j = a_j - \frac{1}{T} \sum_{t=1}^T a_j(t), \quad j \in \{x,y,z\}.
\]

The optimal 1D signal is obtained via projection onto $\mathbf{n}$:
\[
s(t) = \tilde{a}_x(t) n_x + \tilde{a}_y(t) n_y + \tilde{a}_z(t) n_z.
\]

The U-Net v3 model is applied to $s(t)$ to produce the final AO peak set and heart rate estimate.

\begin{table}[th]
\centering
\caption{Experimental Results}
\label{tab:results}
\begin{tabular}{l l cc cccc}
\toprule
Model & Loss Function 
& \multicolumn{2}{c}{Segmentation} 
& \multicolumn{4}{c}{Event-based} \\
\cmidrule(lr){3-4} \cmidrule(lr){5-8}
 &  & IoU (\%) & Dice (\%) 
 & Sensitivity (\%) & PPV (\%) & F1 (\%) & RMSE (ms) \\
\midrule

\multirow{4}{*}{U-Net v3} 
    & BCE          & 93.28 & 96.32 & 99.70 & 99.09 & 99.39 & 18.3 \\
    & Dice + BCE   & \textbf{93.49} & \textbf{96.47} & \textbf{99.80} & \textbf{99.09} & \textbf{99.44} & \textbf{18.5} \\
    & Focal        & 92.85 & 96.10 & 91.48 & 99.45 & 95.30 & 17.6 \\
    & Dice + Focal & 93.06 & 96.19 & 99.49 & 99.09 & 99.29 & 18.6 \\
\midrule

\multirow{4}{*}{U-Net} 
    & BCE          & 89.67 & 94.25 & 97.87 & 98.57 & 98.22 & 22.6 \\
    & Dice + BCE   & 91.51 & 95.34 & 99.49 & 98.89 & 99.19 & 20.7 \\
    & Focal        & 91.14 & 95.14 & 96.04 & 99.27 & 97.63 & 20.2 \\
    & Dice + Focal & 91.31 & 95.20 & 99.29 & 98.89 & 99.09 & 21.8 \\
\midrule

\multirow{4}{*}{U-Net + SA bottleneck} 
    & BCE          & 90.26 & 94.64 & 97.06 & 99.17 & 98.10 & 20.1 \\
    & Dice + BCE   & 90.38 & 94.71 & 97.87 & 98.97 & 98.42 & 20.0 \\
    & Focal        & 89.32 & 94.10 & 84.18 & 99.52 & 91.21 & 20.3 \\
    & Dice + Focal & 90.75 & 94.93 & 98.07 & 99.08 & 98.57 & 19.2 \\
\midrule

\multirow{4}{*}{Patched SACT} 
    & BCE          & 83.86 & 90.84 & 95.64 & 98.64 & 97.12 & 26.7 \\
    & Dice + BCE   & 83.81 & 90.82 & 96.25 & 98.24 & 97.23 & 26.2 \\
    & Focal        & 83.89 & 90.87 & 78.40 & 99.36 & 87.64 & 22.5 \\
    & Dice + Focal & 84.41 & 91.14 & 96.65 & 98.55 & 97.59 & 26.8 \\
\bottomrule
\end{tabular}
\end{table}

\section{Results}
\label{sec:results}

\subsection{Comparative Analysis}
\label{subsec:comparative_analysis}

Table~\ref{tab:results} presents the test-set performance of all evaluated architectures. Event-based metrics are uniformly high ($\geq$95\%), reflecting the quality of post-processing (Section \ref{subsec:postprocessing}). Therefore, meaningful comparison is primarily based on segmentation metrics.

Key findings include:
\begin{enumerate}
    \item U-Net v3 trained with Dice + BCE loss achieved the best segmentation performance, with Dice = 96.47\%, demonstrating near-perfect robustness.
    \item Combining Dice loss with BCE consistently improved sensitivity while maintaining extremely high positive predictive value (PPV), indicating that explicit overlap optimization is beneficial when ground-truth masks are intentionally widened for stable training.
    \item Pure Focal loss systematically reduced sensitivity, particularly in attention- and transformer-based models, suggesting that down-weighting easy negatives is counterproductive in the sparse AO-peak setting.
    \item Classical U-Net architectures, especially the modified U-Net v3, significantly outperformed both the self-attention bottleneck and full transformer architectures (Patched SACT) on segmentation tasks.
\end{enumerate}

In summary, the modernized U-Net v3 trained with Dice+BCE establishes a new state-of-the-art for fully automatic AO detection from smartphone-derived SCG, achieving perfect beat-by-beat detection in our dataset.

\subsection{Model Interpretability}
\label{subsec:interpretability}

\begin{figure}[!t]
    \centering
    \subfloat[Grad-CAM saliency maps (red intensity indicates regions of highest gradient attribution for AO prediction)]{
        \includegraphics[width=1.0\linewidth]{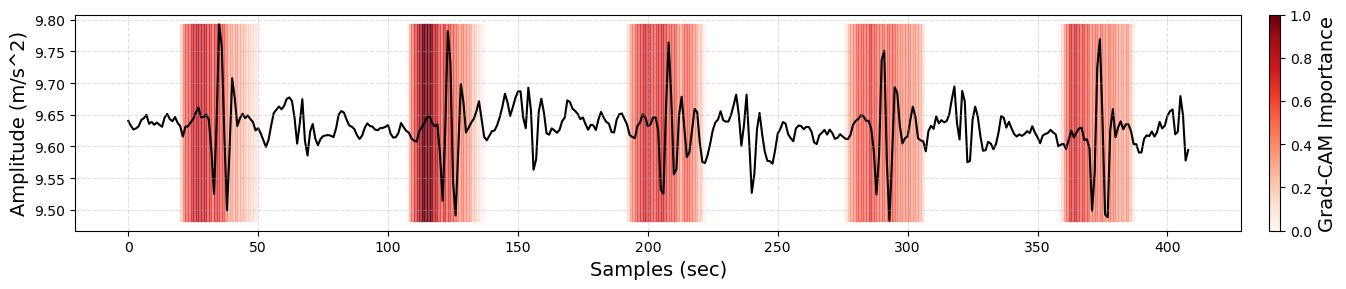}
    }
    \hfill
    \subfloat[Attention-gate activation weights from skip connections (blue intensity reflects feature propagation; darker blue = stronger amplification of AO-related features)]{
        \includegraphics[width=1.0\linewidth]{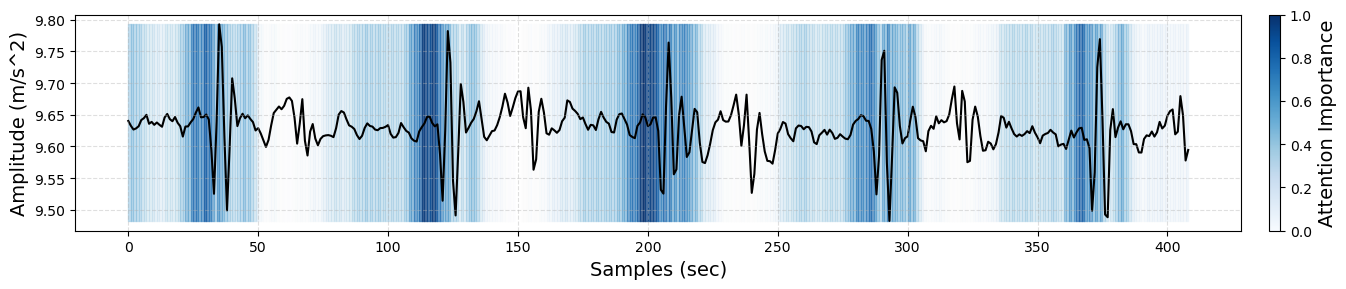}
    }
    \caption{Interpretability analysis of the best-performing model (U-Net v3 + Dice+BCE) on smartphone SCG recordings.}
    \label{fig:interpretability}
\end{figure}

Clinical adoption of deep learning models for cardiac timing requires both trust and interpretability.  

Figure~\ref{fig:interpretability}(a) shows Grad-CAM saliency maps overlaid on the SCG signal. The model consistently assigns the highest importance (dark red) to the exact AO location. Even in segments with significant respiratory modulation and motion artifacts, the gradient attribution remains tightly localized around the true AO instant and largely ignores secondary vibrations and noise.

Figure~\ref{fig:interpretability}(b) visualizes attention weights from the U-Net v3 skip connections. These gates act as learned soft filters, selectively propagating relevant encoder features to the decoder. The attention maps (blue intensity) demonstrate near-perfect alignment with ground-truth AO regions: the network amplifies the characteristic AO waveform while suppressing respiratory components and inter-beat noise. This selective filtering explains the model’s exceptional robustness and near-zero false-positive rate after post-processing.

\subsection{Clinical Validity on Unlabeled Data}
\label{subsec:clinical_validation}

To assess real-world reliability, the best-performing model was applied to an independent set of 1.5 hours of previously unseen, unannotated one-minute SCG recordings. Segments were deemed “analyzable” by cardiologists but were not part of training, validation, or test splits.

Cardiologists blindly evaluated the automatically detected AO timestamps, classifying each 60-second segment as:
\begin{itemize}
    \item \textbf{Satisfactory:} at least 95\% of cycles had clinically correct AO marks,
    \item \textbf{Unsatisfactory:} more than 5\% of AO marks were missed or clearly misplaced.
\end{itemize}

The automated pipeline achieved 94\% satisfactory ratings across all experts, confirming that AO-timing quality is clinically acceptable without manual correction.

\section{Conclusion and Future Work}
\label{sec:conclusion}

We demonstrated that accurate and robust detection of aortic valve opening (AO) events from smartphone-recorded SCG is achievable even in uncontrolled, real-world conditions. By combining a representative dataset of raw triaxial accelerometer recordings, a modified U-Net architecture (U-Net v3) with multi-scale convolutions, residual connections, attention gates, hybrid loss functions, and advanced post-processing, we achieved state-of-the-art performance: sensitivity 99.80\%, PPV 99.09\% across heterogeneous devices and environments.

The stability and consistency of this solution represent a crucial step toward reliable field-deployable cardiac monitoring. Unlike previous SCG studies limited to controlled settings, our approach operates without constraints on body position, phone model, or sensor orientation, enabling truly ubiquitous, long-term, high-frequency rhythm monitoring.

The medical community has already shown interest in the technology. Formal requests for the diagnosis of aortic stenosis have been received, and we have successfully designed and preliminarily validated heuristic features for detecting characteristic morphological changes associated with this pathology in field conditions.

Looking ahead, integrating seismocardiography with large language models (LLMs) and vision-language models (VLMs) presents considerable potential. Multimodal systems could combine raw SCG waveforms, derived mechanical biomarkers, simultaneous photoplethysmography (PPG), and user-reported symptoms to create an end-to-end automated diagnostic pipeline capable of real-time rhythm and pathology classification, interpretable report generation, and interactive dialogue with clinicians or patients.

Ultimately, the transition of seismocardiography from research to a clinical-grade tool is now a realistic near-term goal, enabling population-scale early detection and continuous monitoring of cardiovascular disorders.

\bibliographystyle{IEEEtran}
\bibliography{references}

\end{document}